\documentclass[fleqn,10pt]{wlscirep}
\usepackage[utf8]{inputenc}
\usepackage[T1]{fontenc}
\usepackage{comment}
\usepackage{longtable}
\usepackage{tabularx}
\usepackage{xspace}
\usepackage{float}
\usepackage{soul}
\usepackage{hhline}
\usepackage{csquotes}
\usepackage{xcolor,bytefield}
\usepackage[colorinlistoftodos,prependcaption,textsize=tiny]{todonotes}
\usepackage{ctable}

\definecolor{umn_maroon}{RGB}{122, 0, 25}

\newcommand{\PH}{\ensuremath{\mathrm{H}}\xspace}
\newcommand{\PQb}{\ensuremath{\mathrm{b}}\xspace}

\newcommand{\bbbar}{\ensuremath{\PQb\overline{\PQb}}\xspace}

\newcommand{\pt}{\ensuremath{p_{\mathrm{T}}}\xspace}
\newcommand{\GeV}{\ensuremath{\,\text{Ge\hspace{-.08em}V}}\xspace}
\newcommand{\TeV}{\ensuremath{\,\text{Te\hspace{-.08em}V}}\xspace}

\newcommand{\GEANTfour} {{\textsc{Geant4}}\xspace}
\newcommand{\PYTHIA} {{\textsc{pythia}}\xspace}

\newcommand{\MADGRAPH} {{\textsc{MADGRAPH}}\xspace}
\newcommand{\MCATNLO} {{\textsc{MCATNLO}}\xspace}
\newcommand{\MGvATNLO}{\MADGRAPH{}5\_a\MCATNLO\xspace}
\providecommand{\mSD}{\ensuremath{m_\mathrm{SD}}\xspace} % msd

\title{A FAIR and AI-ready Higgs boson decay dataset}

\author[1,2]{Yifan Chen}
\author[2,3,*]{E. A. Huerta}
\author[4]{Javier Duarte}
\author[5]{Philip Harris}
\author[1]{Daniel S. Katz}
\author[1]{Mark S. Neubauer}
\author[4]{Daniel Diaz}
\author[4]{Farouk Mokhtar}
\author[4,5]{Raghav Kansal}
\author[6]{Sang Eon Park}
\author[1]{Volodymyr V.  Kindratenko}
\author[1]{Zhizhen Zhao}
\author[7]{Roger Rusack}

\affil[1]{University of Illinois at Urbana-Champaign, Urbana, Illinois 61801, USA}
\affil[2]{Argonne National Laboratory, Lemont, Illinois 60439, USA}
\affil[3]{University of Chicago, Chicago, Illinois 60637, USA}
\affil[4]{University of California San Diego, La Jolla, California 92093, USA}
\affil[5]{Hal{\i}c{\i}o\u{g}lu Data Science Institute, La Jolla, California 92093, USA}
\affil[6]{Massachusetts Institute of Technology, Cambridge, Massachusetts 02139, USA}
\affil[7]{The University of Minnesota, Minneapolis, Minnesota, 55405, USA}
\affil[*]{elihu@anl.gov}

\begin{abstract} %170 word limit
To enable the reusability of massive scientific datasets by humans and machines, researchers aim to adhere 
to the principles of findability, accessibility, interoperability, and reusability (FAIR) for data and artificial intelligence 
(AI) models. This article provides a domain-agnostic, step-by-step assessment guide to evaluate whether or not 
a given dataset meets these principles. We demonstrate how to use this guide to evaluate the FAIRness of an 
open simulated dataset produced by the CMS Collaboration at the CERN Large Hadron Collider. 
This dataset consists of Higgs boson decays and quark and gluon background, and is available through the CERN 
Open Data Portal. We use additional available tools to assess the FAIRness of this dataset, and incorporate 
feedback from members of the FAIR community to validate our results. This article is accompanied by a Jupyter 
notebook to visualize and explore this dataset. This study marks the first in a planned series of articles that will 
guide scientists in the creation of FAIR AI models and datasets in high energy particle physics.
\end{abstract}

\begin{document}

\flushbottom
\maketitle

\thispagestyle{empty}

\section*{Introduction}

Much of the success of applications of artificial intelligence (AI) to a broad range of scientific problems~\cite{DeepLearning,Nat_Rev_2019_Huerta} has been due to the availability of well-documented, high-quality datasets~\cite{cvpr:Deng}; open source, state-of-the-art neural network models~\cite{he2016deep,2016wavenet}; highly efficient and parallelizable numerical optimization methods~\cite{pmlr-v28-shamir13}; and the advent of innovative hardware architectures~\cite{Vazquez2013}.

Across science and engineering disciplines, the rate of adoption of AI and modern computing methods has been varied~\cite{Nat_Rev_2019_Huerta}. 
Throughout the process of harnessing AI and advanced computing, researchers have realized that the lack of an agreed upon set of 
best practices to produce, collect, and curate datasets has limited the combination of disparate datasets that with AI may reveal new correlations or patterns~\cite{2020MNRAS.493.3178W,2021arXiv210713049W}.

From 2014 to 2016, a set of data principles, or best practices, based on findability, accessibility, interoperability, and reusability (FAIR) were defined so that scientific datasets could be readily reused by both humans and machines. 
The FAIR principles can be applied to address these limitations and increase the potential of AI for discovery in science and engineering.  
Using high energy physics (HEP) as an example, this article provides  a domain-agnostic, step-by-step set of checks to guide in the process of making a dataset FAIR (``FAIRification").

In HEP, there is a long history of the application of machine learning (ML) techniques to find small signals in the presence of large backgrounds. The observation of the Higgs boson at the CERN Large Hadron Collider (LHC) in 2012~\cite{Aad:2012tfa,Chatrchyan:2012ufa}  was the result of the extensive use of ML algorithms based on boosted decision trees. 
Since then, as ML techniques have developed, their use in HEP has become ubiquitous. However, these developments have been largely the result of physicists adopting AI tools developed outside of their field of research. 

The authors of this paper are members of the \href{https://fair4hep.github.io/team/}{FAIR4HEP collaboration} 
which has representation from the AI community and two of the large LHC collaborations, ATLAS and CMS. 
We are collaborating to prepare datasets from HEP experiments that meet FAIR data principles~\cite{fairguiding}. 
There are several major impediments to this strategy, including, among others, the lack of jargon-free documentation, difficulty of access to, and poor structure of the dataset, and the lack of clear metrics with which to benchmark and compare AI models. 
A consequence of the FAIR data principles is that they promote the use of open datasets, which in turn supports collaboration between practitioners of different disciplines (in this case, high-energy physicists and AI researchers) who have overlapping interests in particular datasets.
We note that these impediments are common to many disciplines.

The FAIR guiding principles~\cite{fairguiding}, published in 2016, provide guidelines to improve the ``FAIRness'' of digital assets, such as datasets, code, and research objects. 
While the principles are valuable for the scientific and engineering fields, they do not include exemplar metrics~\cite{fairmetrics} or define ways to measure how well the FAIR data principles are met for the digital asset. 

In HEP, there are currently several efforts available for creating, indexing, and sharing open, public datasets. 
The \href{http://opendata.cern.ch}{CERN Open Data Portal} provides access to data and resources from the four major LHC collaborations, ALICE, ATLAS, CMS and LHCb, for both education and research. 
Previous data releases have already yielded publications using LHC data authored by external researchers unaffiliated with an LHC collaboration~\cite{Tripathee:2017ybi,Larkoski:2017bvj,Andrews:2018nwy,Andrews:2019faz,Komiske:2019fks,Komiske:2019jim}. 
However, despite being a repository of LHC open data, it does not allow general members of the HEP research community to upload their own datasets.
\href{https://www.zenodo.org}{Zenodo} is another platform launched in May 2013, which is part of the \href{https://www.openaire.eu}{OpenAIRE project}, in partnership with CERN, that is a catch-all repository for European Commission funded research and is used widely. 
It allows the community at large to upload data, software, and other artefacts in support of publications, as well as material associated with conferences, projects, or institutions.
Citation information is also passed to DataCite and other scholarly aggregators. Zenodo has been used to host several high-profile public HEP datasets including the \href{https://doi.org/10.5281/zenodo.2603256}{top quark tagging reference dataset}~\cite{Kasieczka:2019dbj}, \href{https://doi.org/10.5281/zenodo.4536624}{LHC Olympics 2020 
Anomaly Detection Challenge dataset}~\cite{Kasieczka:2021xcg}, \href{https://doi.org/10.5281/zenodo.3601436}{\texttt{hls4ml} jet substructure dataset}, and the \href{https://doi.org/10.5281/zenodo.5070455}{Anomaly Detection Data Challenge 2021 dataset}~\cite{Govorkova:2021hqu}. Other services like Kaggle and Codalab have been used to host HEP challenges like the \href{https://www.kaggle.com/c/trackml-particle-identification}{TrackML accuracy phase}~\cite{Amrouche:2019wmx} and \href{https://competitions.codalab.org/competitions/20112}{TrackML} \href{https://doi.org/10.5281/zenodo.4730157}{throughput phase}~\cite{Amrouche:2021tio}.
The \href{https://www.hepdata.net}{Durham High-Energy Physics Database} (HEPData)~\cite{Maguire:2017ypu} is an open-access repository established for sharing scattering data from experimental particle physics. 
It mainly comprises the data points from plots and tables related to several thousand publications including those from the LHC. 

Despite the widespread availability of public datasets in HEP, these services, and the datasets they host, do not follow FAIR principles. 
In particular, the interpretation of the FAIR principles in the context of the large datasets available and the specific computing infrastructure needs in HEP is not clear. 
For instance, the \href{http://opendata.cern.ch}{CERN Open Data Portal} hosts datasets with sizes approaching 100\,TB~\cite{vbfparked}, requiring special versions of software (provided through a virtual machine image) to read and analyze the data. Given these stringent computational, storage, and domain knowledge requirements, the accessibility of these datasets to non-experts and those lacking resources is not completely obvious. 
To explore how to address these difficulties, we present an analysis of the FAIRness of one of these datasets, the CMS $\PH(\bbbar)$ dataset.

This simulated collider dataset contains a selection of proton-proton interactions (events) in which a Higgs boson is produced and decays to two bottom quarks $\PH(\bbbar)$ (signal events) as well as background events comprised uniquely of ``jets'' of particles produced through the strong interaction, referred to as quantum chromodynamics (QCD) multijet events.
This dataset was released in the \href{http://opendata.cern.ch/}{CERN Open Data Portal}. 
By providing the  details of the how we evaluate FAIRness of this dataset and the steps taken to meet the FAIR data principles, we can help researchers in other fields create FAIR datasets in a similar manner. 
To ensure the reliability of our results we have  conducted a similar study using the \href{https://ardc.edu.au/resources/working-with-data/fair-data/fair-self-assessment-tool/}{ARDC FAIR self assessment tool}. We have found that the steps we have followed and the ARDC assessment tool provide consistent results.

In the following sections we describe how the FAIRness of this dataset is evaluated and present the result in the format of a set of checks. 
We also describe methods to improve FAIRness, provide a detailed data description, and discuss how we interpret FAIR principles for HEP.

\section*{Results}

We have assessed the FAIRness of the target dataset, described in the previous section and in more detail below  using the related FAIR metrics~\cite{fairmetrics}. 
The results of this analysis are summarized in Tables~\ref{tab:find_access} and~\ref{tab:inter_reuse}. 
The following subsections summarize the results for each principle, and discuss steps that were, or will be, take to increase the FAIRness of the dataset. 
We also highlight the difficulties inherent in interpreting and applying these principles to HEP datasets, due to their unique properties of size, complexity, data format, and required domain knowledge.

\subsection*{Findable}
The findable principle requires that metadata and data should be easy to find for both humans and machines. 
For this specific dataset: 1) both data and metadata are registered with globally unique and persistent identifiers;  2) the association between metadata and the dataset itself is explicitly described in its metadata, and 3) the dataset is registered as a searchable resource and is searchable on a commonly used search engine. However, though searchable, the metadata fields are fairly sparse and information is lacking. 
Enriching the metadata with additional fields to include references that cite this dataset, or links to related or derived datasets, would make the data more readily available. 

\subsection*{Accessible}
To meet the FAIR accessible principle, data are required to be kept in a storage facility where they may be easily accessed or downloaded, with well-defined license and access conditions, which should be open access whenever possible, either at the level of metadata, or at the level of the actual data content. 

The CMS $\PH(\bbbar)$ dataset is retrievable using standard HTTP communication protocols, is open access, and is under the \href{https://creativecommons.org/publicdomain/zero/1.0/}{Creative Commons public domain dedication (license)}. 
Since the DOI has formal metadata, it satisfies the metadata longevity plan.

\subsection*{Interoperable}
The interoperable principle requires that the data can be readily combined with other datasets by humans as well as by computer systems. 
For this dataset, (meta)data are represented using a formal and broadly applicable representation language. 
To improve the interoperability, the data descriptions were rewritten to be human-readable, removing jargon to make it accessible not only for domain experts, but also non-HEP researchers. The (meta)data use a set of FAIR vocabularies defined for both general purpose and HEP domain-related purpose. 
Although not all terms are findable in FAIR vocabularies, those that are not findable are well-defined and referenced. Lastly, the description of this dataset provides references to other datasets from which it is derived.
However, a more extensive set of references that elaborate on the paper describing this dataset, and more information about the methods used to derive this dataset, could be added to aid in the comprehension of the problem to be addressed with this dataset.

\subsection*{Reusable}
Reusability requires the data to be readily usable for future research and to be able to be processed further using different computational methods. 
We found that the metadata and data of this dataset are well-described with accurate and relevant attributes. 
Thus, we anticipate that the dataset will be reusable and can be integrated with additional data in future studies.

\newpage

\begin{longtable}[h]{|p{6cm}|p{10.5cm}|}
\caption{\textbf{Findable} and \textbf{Accessible} principle assessment checks for the CMS H($b\bar{b}$) Open Dataset.
\label{tab:find_access}}\\
\hline
\emph{Metric} & \emph{Evaluation}\\
\hline
%F1
\multicolumn{2}{|p{14cm}|}{
F1. (Meta)data are assigned globally unique and persistent identifiers.} \\
\hline 
\textbf{Identifier Uniqueness}: this metric measures whether there is a scheme to uniquely identify the digital resource. 
& 
\textbf{Pass}. The  \href{http://doi.org/10.7483/OPENDATA.CMS.JGJX.MS7Q}{DOI}
for the data (which resolves to a URL~\cite{ref:dataset}) follows a registered identifier scheme. \\
\hline
\textbf{Identifier Persistence}: this measures whether there is a policy that describes what the provider will do in the event an identifier scheme becomes deprecated. & \textbf{Pass}. The use of a DOI provide a persistent interoperable identifier. \newline \\
\hline
% F2
\multicolumn{2}{|p{14cm}|}{F2. Data are described with rich metadata.} \\ 
\hline 
\textbf{Machine-readability of Metadata}: to meet this metric, a URL to a document containing machine-readable metadata for the digital resource must be provided. 
&
\textbf{Pass.} The URL for the metadata~\cite{ref:metadata} in \href{https://json-schema.org/}{JSON Schema} 
with REST API is available. The use of JSON Schema provides clear human and machine readable documentation. 
Also, running the URL through the \href{https://search.google.com/test/rich-results}{Rich Result Test} shows the data page contains rich results.\\
\hline
\textbf{Richness of Metadata}: data are described with rich metadata
&
\textbf{Partially pass.} 
Reviewing the \href{https://ez.datacite.org/id/doi:10.7483/OPENDATA.CMS.JGJX.MS7Q}{DataCite metadata} 
for the DOI shows a fairly sparse record. The metadata can be improved with richer fields. \\
\hline
% F3 
\multicolumn{2}{|p{14cm}|}{F3. Metadata clearly and explicitly include the identifier of the data they describe.} \\ 
\hline 
\textbf{Resource Identifier in Metadata}: this measures if the metadata document contains the identifier for the digital resource that meets F1 principle.
&
\textbf{Pass.} The association between the metadata and the dataset is made explicit because the dataset's globally unique and persistent identifier can be found in the metadata. 
Specifically, the DOI is a top-level and a mandatory field in the metadata record. \\
\hline
% F4
\multicolumn{2}{|p{14cm}|}{F4. (Meta)data are registered or indexed in a searchable resource}\\ 
\hline
% \emph{Metric} & \emph{Evaluation} \\
% \hline
\textbf{Index in a searchable resource}: this measures the degree to which the digital resource can be found using web-based search engines 
&
\textbf{Pass.} The dataset is indexed by Google Dataset Search engine. \\
\hhline{|==|}
% Accessible
\multicolumn{2}{|p{14cm}|}{
A1. (Meta)data are retrievable by their identifier using a standardized communications protocol \newline 
A1.1: The protocol is open, free and universally implementable}\\
\hline
\textbf{Access Protocol}: it measures whether the URL is open access and free.
&
\textbf{Pass}. HTTP get on the identifier's URL returns a valid document \\
\hline
%A1.2
\multicolumn{2}{|p{14cm}|}{
A1.2. The protocol allows for an authentication and authorization where necessary}\\
\hline
\textbf{Access Authorization}: it requires specification of a protocol to access restricted content.
&
\textbf{Pass}. This is an open dataset, accessible to everyone on the internet. 
The data is non-profit and privacy-unrelated, so no access authorization is needed.\\
\hline
% A2
\multicolumn{2}{|p{14cm}|}{
A2. Metadata should be accessible even when the data is no longer available}\\
\hline
\textbf{Metadata Longevity}: it requires metadata to be present even in the absence of data
&
\textbf{Pass.} Metadata is stored separately in the CERN Open Data server. 
As per FAIR Principle F3, this metadata remains discoverable, even in the absence of the data, because it contains an explicit reference to the DOI of the data. Data and metadata will be retained for the lifetime of the repository. 
The host laboratory CERN,  currently plans to support the repository for at least the next 20 years. \\
\hline
\end{longtable}
\newpage

\begin{longtable}[h]{|p{6cm}|p{10.5cm}|}
\caption{\textbf{Interoperable} and \textbf{Reusable} principle assessment checks for CMS H($b\bar{b}$) Open Dataset\label{tab:inter_reuse}}\\
\hline
\emph{Metric} & \emph{Evaluation} \\
\hline
%I1
\multicolumn{2}{|p{14cm}|}{
I1. (Meta)data use a formal, accessible, shared, and broadly applicable language for knowledge representation.}\\
\hline
\textbf{Use a Knowledge Representation (programming) Language}: use a formal, accessible, shared, and broadly applicable language for knowledge representation
&
\textbf{Pass.} As described in Section 3, this dataset is represented based on the \texttt{ROOT} framework with Python interface. The notebook we release with this manuscript provides the required tools to handle this dataset using \texttt{HDF5}. 
The metadata is represented following the \href{https://json-schema.org/specification-links.html#draft-4}{JSON Schema draft 4}. Both are widely used formats in Physics.  \\

\hline
\textbf{Provide Human-readable descriptions} & \textbf{Pass}. The description and data semantics of this dataset provides rich information on how to use the dataset.  \\
\hline
% I2
\multicolumn{2}{|p{14cm}|}{
I2. (Meta)data use vocabularies that follow FAIR principles.} \\
\hline
\textbf{Use FAIR Vocabularies}: it requires the metadata values and qualified relations should be FAIR themselves, that is, terms should be findable from open, community-accepted vocabularies.
& 
\textbf{Partially pass}. I2 requires the controlled vocabulary used to describe datasets to be documented and resolvable using globally unique and persistent identifiers. 
For domain-specific terms, we leverage a vocabulary PhySH (\href{https://physh.org/contribute}{Physics Subject Headings}), a physics classification scheme developed by American Physical Society (APS). 
Some terms in dataset descriptions and semantics are registered in PhySH. 
However, since PhySH is still under development, there is not very good coverage of the narrower experimental concepts. 
For the terms not covered, references and hover definitions are provided. 
For general terms, the metadata follows the vocabulary from \href{https://json-schema.org/}{JSON Schema} and a minimal set of FAIR terms are used. \\

\hline
% I3
\multicolumn{2}{|p{14cm}|}{
I3. (Meta)data include qualified references to other (meta)data.}\\
\hline
\textbf{Use Qualified References}: The goal is to create as many meaningful links as possible between (meta)data resources to enrich the contextual knowledge about the data. 
&
\textbf{Partially pass.} There are connections with other datasets. 
A list of derived datasets is available at the dataset site \cite{ref:dataset}.
Each referenced external piece of dataset is qualified by a resolvable URL and a unique CERN data identifier in metadata. 
To improve, the papers of these related data can be provided, from which more information about methods and workflow used to derive this dataset can be retrieved, and external datasets should be references by permanent identifiers rather than URLs. \\
\hhline{|==|}
\multicolumn{2}{|p{14cm}|}{R1.1.  (Meta)data are released with a clear and accessible data usage license.} \\
\hline
\textbf{Accessible Usage License}: the existence of license document for (meta)data are being measured 
 &
\textbf{Pass.} This dataset is released under \href{https://creativecommons.org/publicdomain/zero/1.0/}{Creative Commons CC0 dedication}. The license field is present in the metadata.\\
\hline
\multicolumn{2}{|p{14cm}|}{R1.2. (Meta)data are associated with detailed provenance.} \\
\hline
\textbf{Detailed Provenance}: Who / What / When produced the data? Why / How was the data produced?
&  \textbf{Pass.}  The dataset is derived from other data, e.g.~\cite{BulkGravTohhTohbbhbb_narrow_M-600,QCD_Pt_300to470}, using public software~\cite{ntupler} that was made public to process and reduce it. 
We are able to track the original authors and data sources.
But ideally, this workflow would be described in a machine-readable format. \\
\hline
\multicolumn{2}{|p{14cm}|}{
R1.3. (Meta)data meet domain-relevant community standards.} \\
\hline
\textbf{Meet Community Standards}: it measures whether a certification of the resource meeting community standards exists. 
&
\textbf{Pass.} Both metadata and data meet the CERN Open Data community standards and thus have been released on the CERN Open Data repository. \\
\hline
\end{longtable}

\section*{Methods}

In this section we describe our approach to evaluate FAIRness of the CMS $\PH(\bbbar)$ dataset, and provide a human-readable description of the HEP dataset contents and its overall structure. 
These two complementary aspects of the dataset are critical elements in any pursuit of data FAIRification. 

\subsection*{Dataset FAIRification}
\label{sec:data_fair}

We have created a set of ready-to-use, domain-agnostic checks to facilitate the evaluation of how well a dataset meets the FAIR guiding principles~\cite{fairguiding}, and applied them them to the $\PH(\bbbar)$ dataset. 
These checks provide researchers with a tool that can be used to assess the FAIRness of scientific datasets, and thus will streamline the use of such datasets for AI-driven analyses. 

We have used the \href{https://ardc.edu.au/resources/working-with-data/fair-data/fair-self-assessment-tool/}{ARDC FAIR self assessment tools}, developed by other researchers in the FAIR community, to validate our findings. 
We have also incorporated human-in-the-loop expertise in this process in the form of feedback from FAIR experts, who independently validated our results.

\subsection*{Dataset Description}
\label{sec:data}

The CMS $\PH(\bbbar)$ Open Dataset consists of two data samples that have been a critical part of the understanding of physical phenomena associated with the Higgs boson. 
The Higgs boson, first observed at the LHC in 2012~\cite{Aad:2012tfa,Chatrchyan:2012ufa}, is an elementary particle that is related to the Higgs mechanism for electroweak symmetry breaking, responsible generating the masses of the elementary particles.

One consequence of the Higgs mechanism is that the Higgs boson, which has a lifetime of only $\approx 10^{-22}$ seconds, couples to other particles in proportion to their mass and therefore will decay preferentially to elementary particles with comparatively higher masses. The $\PH(\bbbar)$ decay process is particularly important because the $\PQb$ quark is the most massive quark to which the Higgs boson can decay. 
By measuring precisely the rate of this decay process, the physics of the coupling between the Higgs boson and ordinary matter can be tested. Any significant deviations from the predicted values would be an indication of physics beyond the standard model of particle physics.

When a Higgs boson decays to $\PQb$ quarks, the quarks, which cannot be free in nature, are detected as clusters of particles moving away from the interaction vertex (jets) and recognized by a secondary decay vertex from a particle containing a $\PQb$ quark a short distance from the interaction. 
Collisions, or interactions between protons in the two circulating beams (events) occur at a rate of about 1~GHz, while the rate of production of Higgs bosons is only 0.001~Hz, about one every hour. 
The challenge of identifying Higgs bosons decaying to $\bbbar$ is to find them amid the much larger number of collisions (background) where a Higgs boson is not produced. 
In these background events, typically referred to as quantum chromodynamics (QCD) multijet events, a large number of particles are produced, which may include jets from  $\PQb$ quarks, and can combine to resemble $\PH(\bbbar)$ events, which are the ``signal'' events.

To identify Higgs boson decays and separate them from the much larger QCD background, we use several key reconstructed components of proton-proton collisions. 
In particular, we reconstruct jets and analyze their characteristics which include tracks, secondary vertices (SVs), and substructure features. 
We also employ a particle-flow (PF) algorithm~\cite{CMS-PRF-14-001} to provide a comprehensive list of final-state particles that are identified and reconstructed via combination of information from multiple detector subsystems.

The following defines these elements: 
\begin{itemize}
    \item {\bf Jets} are sprays of elementary particles in a cone-shaped pattern that radiate out from the collision vertex.
    They may be characterized by their \emph{substructure}, including features like the jet mass, charge, and shape~\cite{Thaler:2010tr}.
    In total, the dataset contains 64 reconstructed jet features. 
    These features are not necessarily independent from one another, and they may be derived from lower-level features related to the tracks, PF candidates, and secondary vertices.
    \item {\bf Tracks} are the reconstructed helical paths of charged particles as they move away from the collision vertex in the magnetic field at the detector.  
    Each charged particle leaves a characteristic set of hits in the tracking detector of CMS, which are used to reconstruct the track.
    In total, there are 45 track features.    
    \item {\bf Secondary vertices} are collections of tracks that originate from a particle decay that is not at the collision vertex. 
    Secondary vertices are a interesting set of candidate features that discriminate between different classes of jets, because they are dominant signatures in bottom-quark decays.
    In total, there are 18 SV features.
    \item {\bf Particle-flow candidates} are formed by combining tracks and clusters from other detectors outside of the tracking detector.
    The PF algorithm~\cite{CMS-PRF-14-001} is used to provide a complete event description through the generation of a comprehensive list of the particles produced in the collision. 
    For each PF candidate, there are 24 features.
\end{itemize}

This dataset consists of particle jets extracted from simulated proton-proton collision events generated with a center-of-mass energy of 13\TeV. 
Each element of the dataset corresponds to a single jet, containing information about the jet, from jet-level features to track-level features (see later for the full details on the dataset). 

The outcome of the default CMS reconstruction workflow is provided in the open simulation~\cite{ref:dataset}. In particular, particle candidates are reconstructed using the particle-flow (PF) algorithm~\cite{CMS-PRF-14-001}. 
Particles produced nearly simultaneously with the events that leave extra hits in the detector (pileup) are removed with an algorithm developed for that purpose~\cite{CMS:2020ebo}.
Jets are clustered from the remaining reconstructed particles~\cite{Cacciari:2008gp, Cacciari:2011ma} with a jet-size parameter $R=0.8$ (AK8 jets). 
The standard CMS jet energy corrections are applied to the jets. In order to remove soft, wide-angle radiation from the jet, the soft-drop (SD) algorithm~\cite{Dasgupta:2013ihk,Butterworth:2008iy} is applied, with  angular exponent $\beta = 0$, soft cutoff threshold $z_{\mathrm{cut}} < 0.1$, and characteristic radius $R_{0} = 0.8$~\cite{Larkoski:2014wba}.
The SD mass ($\mSD$) is then computed from the four-momenta of the remaining constituents.

The dataset is reduced by requiring the AK8 jets to have $300 < \pt < 2400 \GeV$, $|\eta| < 2.4$, and $40 < \mSD < 200 \GeV$. 
After this reduction, the dataset consists of 3.9 million $\PH(\bbbar)$ jets and 1.9 million QCD jets. Charged particles are required to have $\pt > 0.95 \GeV$ and reconstructed secondary vertices (SVs) are associated with the AK8 jet using $\Delta R = \sqrt{\Delta\phi^2+\Delta\eta^2} < 0.8$.
The dataset is divided into blocks of features, referring to different objects: tracks, secondary vertices, and particle candidates. See the \href{http://opendata.cern.ch}{CERN Open Data Portal}  for a complete list of features.

\begin{figure}[htbp]
    \centering
    \includegraphics[width=0.49\textwidth]{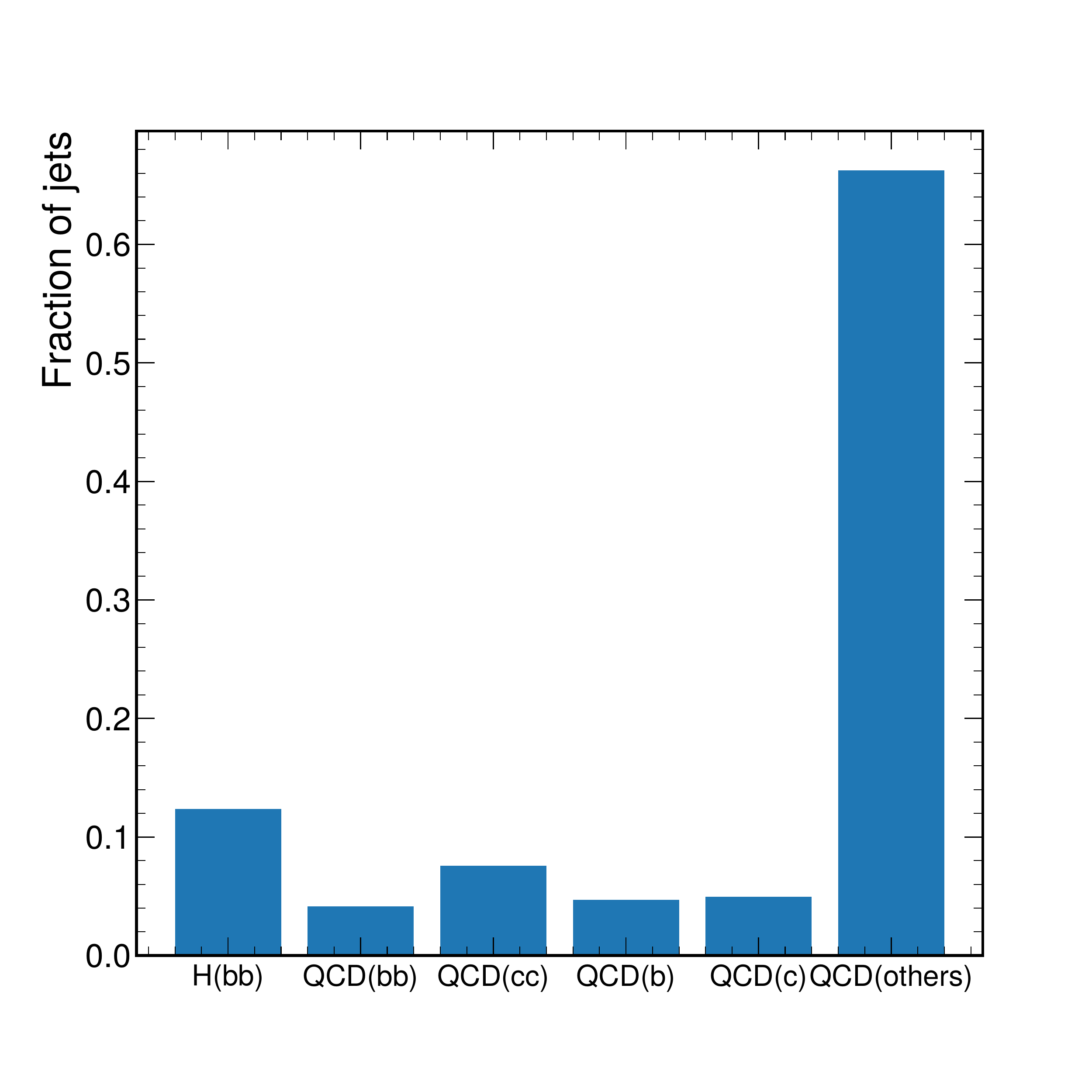}
    \caption{The distribution of labels is shown for a representative file in the training dataset.}
    \label{fig:labels}
\end{figure}

A typical use case for this dataset is the development of a machine-learning classifier to distinguish the $\PH(\bbbar)$ signal from the QCD background jets, which in ML terms, can be done via a binary-classification task. 
However, it is often useful to further classify the QCD jets, thereby, the task becomes multi-class classification with the following six jet classes: \texttt{H\_bb}, \texttt{QCD\_bb}, \texttt{QCD\_cc}, \texttt{QCD\_b}, \texttt{QCD\_c}, and \texttt{QCD\_others}. 
The labeling is performed sequentially. 
If a `generator-level' Higgs boson is geometrically matched to the AK8 jet ($\Delta R < 0.8$) and the two bottom quark decay products are also matched to the jet, then it is labeled as \texttt{H\_bb}. 
If instead, only two bottom (charm) quarks are found, the jet is labeled as \texttt{QCD\_bb} (\texttt{QCD\_cc}). If only a single bottom (charm) quark is found, it is labeled as \texttt{QCD\_b} (\texttt{QCD\_c}).
Finally, if none of the above conditions are met, it is labeled as \texttt{QCD\_others}. The distribution of labels is shown in Fig.~\ref{fig:labels}. 
The large class imbalance is a common feature of classification problems in high energy physics: background jets occur at much larger rates than signal jets.

\begin{figure}[ht]
    \centering
    \includegraphics[width=0.7\textwidth,viewport=400 200 1520 880,clip=true]{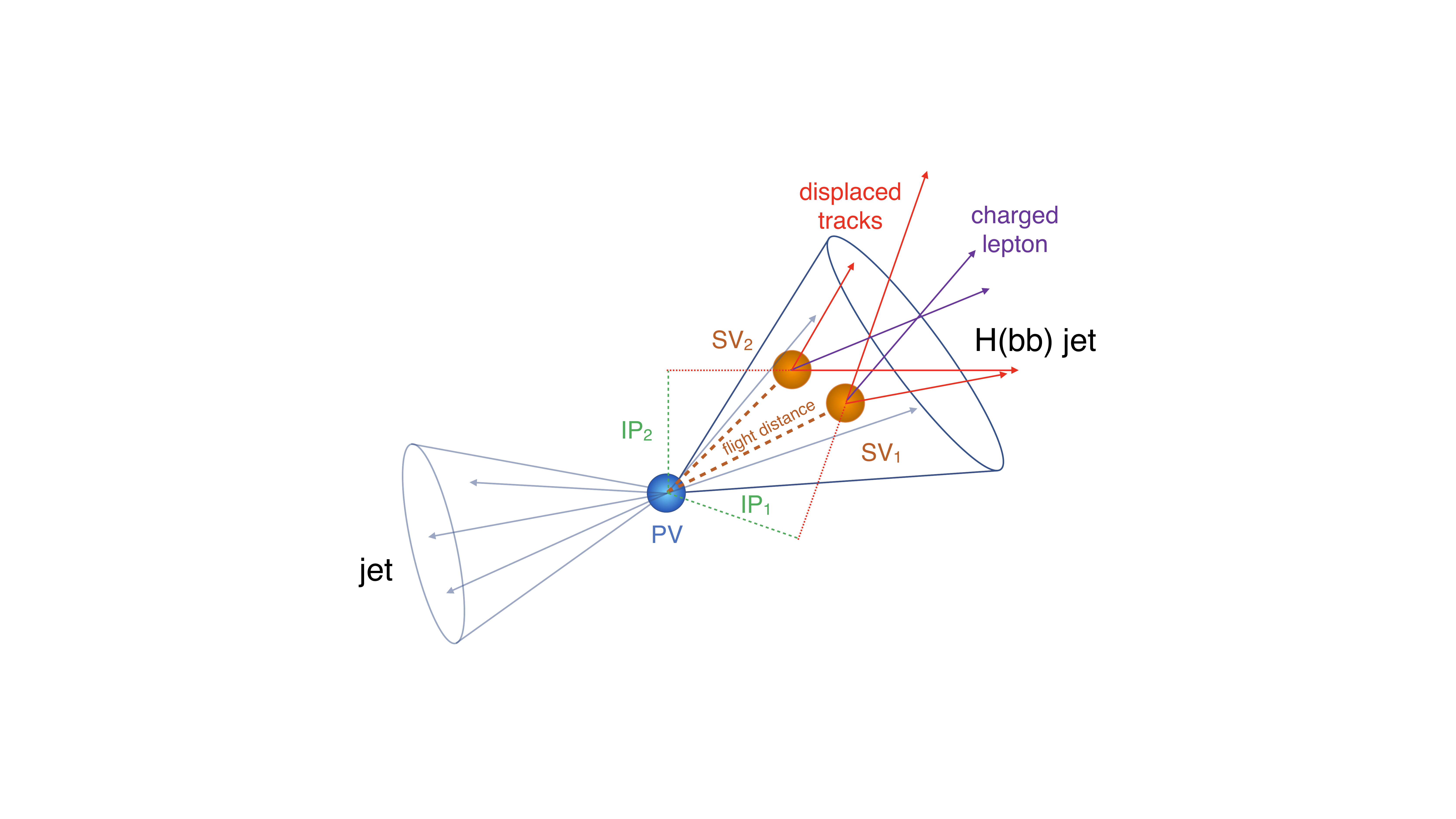}
    \caption{Illustration of a $\PH(\bbbar)$ jet with two secondary vertices (SVs) from the decay of two b hadrons resulting in charged-particle tracks (including a low-energy, or soft, lepton) that are displaced with respect to the primary collision vertex (PV), and hence with a large impact parameter (IP) value.}
    \label{fig:HbbTagging}
\end{figure}

\begin{figure}[htbp]
    \centering
    \includegraphics[width=0.98\textwidth]{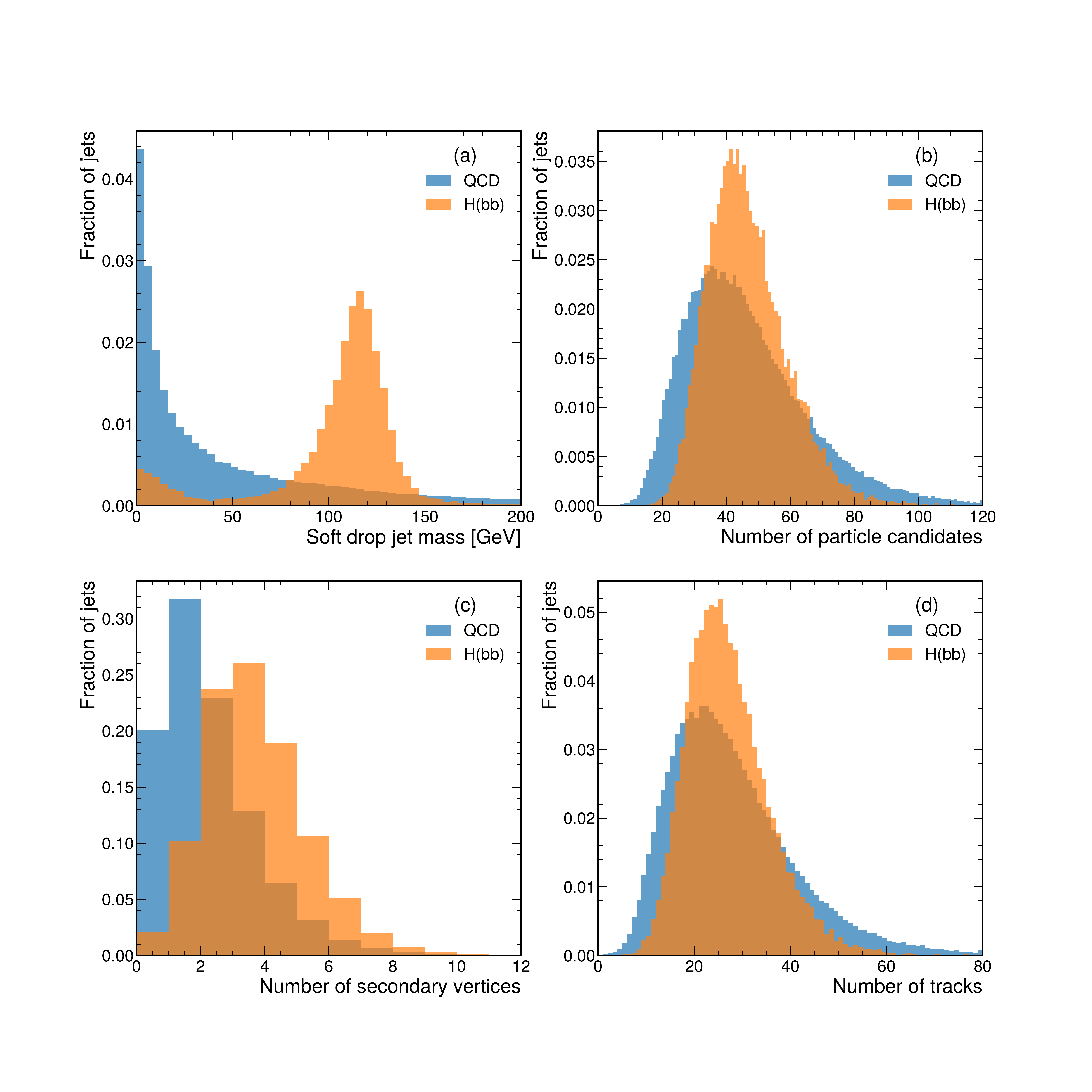}
    \caption{The distributions of some salient jet features: \textbf{(a)} the soft-drop jet mass; \textbf{(b)} number of particle candidates; \textbf{(c)} number of secondary vertices; and \textbf{(d)} number of tracks, are shown for one file in the training dataset.}
    \label{fig:features}
\end{figure}

Specific signatures of \PQb quark decays can be used in a ML algorithm to differentiate between $\PH(\bbbar)$ and QCD jets. 
For instance, one of the distinct signatures of \PQb quarks is its long lifetime, which in a high energy collision translates to a particle that decays with a displacement with respect to the collision. 
The model can learn this information to improve the accuracy of the inference. 
An illustration of some of key features that can be used for $\PH(\bbbar)$ jet tagging are shown in Fig.~\ref{fig:HbbTagging}. The distributions of some salient jet features are shown in Fig.~\ref{fig:features}.

Many different deep learning architectures have been developed and studied for the task of jet classification, such as: interaction networks (INs)~\cite{Battaglia:2016jem}, dynamic graph convolutional neural networks~\cite{Qu:2019gqs}, and Lorentz-group equivariant networks~\cite{Bogatskiy:2020tje}. The first was applied to this data~\cite{Moreno:2019neq} as a comparison with another ML model called the deep double-b (DDB) tagger created by the CMS Collaboration~\cite{CMS-DP-2018-058} that uses a smaller subset of the input features. 
In addition to jet classification~\cite{Sirunyan:2017ezt,Chatrchyan:2012jua,Bols:2020bkb,CMS-DP-2018-058}, a further challenge within this dataset is a regression task, whereby one attempts to reconstruct the true energy of the Higgs boson. 
To perform this task, a regression loss needs to be constructed targeting the true Higgs boson energy. This promotes the exploration of physics-motivated loss functions, such as the earth (or energy) mover's distance (EMD)~\cite{Komiske:2019fks}.

\subsection*{Dataset Structure}
\label{sec:data_structure}

Particle physics uses a variety of data formats (and analysis ecosystems), including the \texttt{ROOT} library~\cite{root}. 
\texttt{ROOT} is a framework for data processing, created at CERN that is widely used by the high-energy physics community. 
A \texttt{ROOT} file is a compressed binary file where objects of any type can be saved. There are Python bindings built into \texttt{ROOT}, which are called \texttt{PyROOT}. 
Recently, an additional library called uproot~\cite{uproot} has been developed that allows Python users to perform \texttt{ROOT} I/O directly.
Unlike the standard \texttt{C++ ROOT} implementation, uproot is only an I/O library, primarily intended to stream data into machine learning libraries in Python. 
It can also make jagged or awkward arrays~\cite{awkward:2020}.

Trees are a data structure in \texttt{ROOT} that are tables of information. 
Trees are composed of \emph{branches}, which are the columns of the table. 
In this dataset, each row represents a jet. 
Some branches contain only a single floating point number per entry (jet). 
Other branches contain a vector of floating point numbers, where the length of the vector varies for each entry.
The former means there is only one number per jet (or event); the latter means there may be a variable number per jet.

In addition to the \texttt{ROOT} format, this dataset is also released in \texttt{HDF5} format. 
The \texttt{HDF5} files contain different arrays for each output variable, with only information for up to 100 particle candidates, 60 tracks, and 5 secondary vertices stored in zero-padded arrays.

\section*{Discussion}
The motivation of our work to adopt FAIR principles for the production, collection and curation of scientific datasets is to streamline and facilitate their use in the design, training, validation and testing of AI models.
This approach is particularly relevant for ongoing efforts that aim to automate the inference of massive scientific datasets through the convergence of AI and modern computing environments~\cite{2021NatAs.tmp..118H,Huerta2020}. 
It is often the case that AI models are trained with (abundant and easy to produce) synthetic data, large scale simulations, and first-principles mathematical models, although these may only provide an incomplete description of complex and highly nonlinear real-world phenomena.
Thus, when AI models are used to extract new knowledge from realistic, experimental datasets, it is a common occurrence that AI predictions are off-target. 
However, once AI models are calibrated against experimental data, their predictions become increasingly accurate~\cite{deepdrive}. 
Given that this is a trend reported across many disciplines, it is useful to streamline the development of AI models with real, experimental datasets. 
This can be accomplished if synthetic and experimental datasets are produced, collected and curated following a common set of standards or, in this case, FAIR guiding principles. 

Another motivation to understand and adopt FAIR principles to create AI-ready datasets is that some disciplines are subject to restrictive regulations that prevent data fusion and centralized analyses. This is a common issue in multi-modal biomedical datasets that are governed by federal regulations, consortium-specific data usage agreements, and institutional review boards. These restrictions have catalyzed the development of federated learning approaches, and the development of privacy-preserving methods and the use of secure data enclaves. 
It is clear that developing AI models by harnessing disparate data enclaves will only be feasible if datasets adhere to a common set of rules, or FAIR principles.

In this study we have shown that open source datasets may not be FAIR or AI-ready. The domain-agnostic checks that we provide in this article will provide researchers with a starting point, and guidance to FAIRify their datasets. The FAIR principles are comprehensive and can be used by AI and domain experts to enable the reusability of massive scientific datasets that will enable the creation of next-generation AI models leading to a digitally accurate, interpretable and reproducible description of natural phenomena. 

Researchers who create FAIR datasets should keep in mind that this work aims to automate end-to-end AI studies, from data collection to inference. This will only be accomplished if datasets contain all the information needed to interpret, verify, and reproduce new findings. 
To accomplish this goal, we recommend that datasets are stored using formats that are widely available in modern computing environments, such as \texttt{HDF5} or \texttt{ROOT}. Using such data formats simplify the handling of large datasets, will allow experimental and synthetic datasets to be on the same footing, and will make accessible more datasets for widely used APIs for AI research, e.g., \texttt{TensorFlow} or \texttt{PyTorch}. 
In future work, we plan to introduce tools to automate the evaluation of FAIR metrics for datasets and to gain a better understanding of the relationship between data and AI models.

\section*{Data Availability} 
The $\PH(\bbbar)$ data for this work is available on the \href{http://opendata.cern.ch}{CERN Open Data Portal}~\cite{ref:dataset}, in both \texttt{ROOT} and \texttt{HDF5} formats.

\section*{Data Formatting} 
The technical details of the simulation of the events and their selection for inclusion in the dataset are described in this section. The dataset consists of a signal model containing $\PH(\bbbar)$ jets available from simulated events containing the postulated Randall-Sundrum gravitons~\cite{Randall} that decay to two Higgs bosons, and thence to $\bbbar$ pairs.

The event generation was done by the CMS Collaboration with \MGvATNLO 2.2.2 at leading order, with graviton masses ranging between 0.6 and 4.5\TeV. Generation of this process enables better sampling of events 
where the Higgs boson is produced with a large lateral momentum component ($\pt$). The background dataset was generated with \PYTHIA 8.205~\cite{Sjostrand:2014zea} in different bins of the average $\pt$ of the final-state partons ($\hat{p}_\mathrm{T}$). The parton showering and hadronization was also performed with \PYTHIA8.205, using the CMS underlying event tune CUETP8M1~\cite{Khachatryan:2015pea} and the NNPDF~2.3~\cite{NNPDF2} parton distribution functions. Pileup interactions are modeled by overlaying each simulated event with additional minimum bias collisions, also generated with \PYTHIA8.205.
The CMS detector response is modeled by \GEANTfour~\cite{Agostinelli:2002hh}.

Data are composed of a set of characteristic variables relating to several broad types of objects consisting of event-level identifiers, features of charged particle tracks, secondary vertices, and particle-flow candidates, high level jet observables, and additional generator-level information, such as jet labels. There are 3 event-level features, 45 charged particle features, 18 secondary vertex features, 24 particle-flow candidate features, 64 high-level jet features, and 18 generator-level identifiers. For each of these variables, a detailed description is present on the \href{http://opendata.cern.ch}{CERN Open Data Portal}. For the \texttt{HDF5} format, information is stored for up to 100 particle-flow candidates, with a maximum of 60 charged particle tracks, and up to 5 secondary vertices. 
In the instance where there are less candidates, inputs are zero-padded.

\section*{Code availability}
To make the CMS $\PH(\bbbar)$ Open Dataset more accessible, we provide notebooks~\cite{UCSD-ParticalML} from the course ``Particle Physics and Machine Learning'' at University of California San Diego. The course notebooks provide a guide for the use of the \texttt{ROOT} format dataset. We also have released a second set of interactive Jupyter Notebooks on GitHub~\cite{binder-notebooks}, where we visualize feature distributions and feature correlations, and provide machine learning examples on low-level features in this dataset. The Jupyter notebooks that we released show how the \texttt{HDF5}-formatted data can be accessed.

\bibliography{sample}

\section*{Acknowledgements} %(not compulsory)

We thank Tom Honeyman from the Australian Research Data Commons (ARDC) and Chris Erdmann 
from the American Geophysical Union (AGU) for their help and advice on both FAIR data principles in 
general and on their application to our specific dataset, though any errors in interpretation of the 
principles are ours. We thank the CMS Collaboration for making the $\PH(\bbbar)$ dataset publicly 
available and for helpful discussions in the preparation of this work. We also thank Tibor Simko, 
Kati Lassila-Perini, and the rest of CERN Open Data Portal Team. This work was performed as 
part of the FAIR Framework for Physics-Inspired Artificial Intelligence in High Energy 
Physics (FAIR4HEP) project (DE-SC0021258, DE-SC0021395, DE-SC0021225, and DE-SC0021396), 
support by the Office of Advanced Scientific Computing Research within U.S. Department of 
Energy Office of Science. FM was partially supported by an Hal{\i}c{\i}o\u{g}lu Data Science Fellowship.

\section*{Author contributions statement}

E.A.H. led this work and coordinated the writing of this manuscript. 
Y.C. and J.D. participated in the selection and FAIRification of our sample dataset. D.S.K. provided 
FAIR expertise to guide the initial FAIR assessment, and then worked with external FAIR experts to 
validate our results. V.K. contributed to the evaluation of the results. M.S.N. and P.H. contributed to the 
FAIR assessment of our sample dataset and as an internal editor of the manuscript. Z.Z. contributed to 
the evaluation of the results and reviewed the manuscript. R.K., F.M., and D.D. provided feedback on the 
manuscript and contributed to the FAIRification of the dataset. S.E.P. provided feedback on the manuscript. 
R.R. acted as an internal editor of the manuscript. All authors reviewed the manuscript. 

\section*{Competing interests} %(mandatory statement)
The authors declare no competing interests. 

\end{document}